\documentclass[12pt]{article}

\usepackage{graphicx,array}
\usepackage{amssymb,amsmath}
\usepackage[english]{babel}

\newcommand\ba{\begin{eqnarray}}
\newcommand\ea{\end{eqnarray}}
\newcommand\be{\begin{equation}}
\newcommand\ee{\end{equation}}

\begin{document}

\begin{titlepage}

\begin{center}

{\Large {\bf Exceptional quantum algebra for the \\
standard model of particle physics\footnote{Extended version of talks at the: 13-th International Workshop "Lie Theory and Its Applications in Physics" (LT13), Varna, Bulgaria, June 2019; Workshop "Geometry and mathematical physics" (to the memory of Vasil Tsanov), Sofia, July 2019; Simplicity III Workshop  at Perimeter Institute, Canada, September 9-12, 2019; Humboldt Kolleg "Frontiers in Physics, from Electroweak to Planck scales", Corfu, September 15-19, 2019; Conference "Noncommutative  Geometry and the Standard Model", Jagiellonian University, Krakow, 8-9 November 2019.

To be published in the Springer Proceedings in Mathematics and Statistics.}}}

\vspace{4mm}

Ivan Todorov

Institute for Nuclear Research and Nuclear Energy\\
Bulgarian Academy of Sciences \\
Tsarigradsko Chaussee 72, BG-1784 Sofia, Bulgaria\\
e-mail: ivbortodorov@gmail.com

\end{center}

\begin{abstract}

The exceptional euclidean Jordan algebra $J_3^8$ of $3\times 3$ hermitian octonionic matrices, appears to be tailor made for  the internal space of the three generations of quarks and leptons. The maximal rank subgroup of the automorphism group $F_4$ of   $J_3^8$ that respects the lepton-quark splitting is $(SU(3)_c\times SU(3)_{ew})/\mathbb{Z}_3$. Its restriction to the special Jordan subalgebra $J_2^8\subset J_3^8$, associated with a single generation of fundamental fermions, is precisely the symmetry group $S(U(3)\times U(2))$ of the Standard Model. The Euclidean extension 
$\mathcal{H}_{16}(\mathbb{C})\otimes \mathcal{H}_{16}(\mathbb{C})$ of $J_2^8$, the subalgebra of hermitian matrices of the complexification of the associative envelope of  $J_2^8$, involves 32 primitive idempotents giving the states of the first generation fermions. The triality relating left and right $Spin(8)$ spinors to 8-vectors corresponds to the Yukawa coupling of the Higgs boson to quarks and leptons.

\smallskip
 
The present study of $J_3^8$ originated in the paper arXiv:1604.01247v2 by Michel Dubois-Violette. It reviews and develops ongoing work with him and with Svetla Drenska: 1806.09450; 1805.06739v2; 1808.08110.  

\end{abstract}

\vfill\eject

\tableofcontents

\end{titlepage}

\section{Motivation. Alternative approaches}
\setcounter{equation}{0}
\renewcommand\theequation{\thesection.\arabic{equation}}
The gauge group of the Standard Model (SM),   
\begin{equation}
\label{GSM} 
G_{SM} = \frac{SU(3)\times SU(2)\times U(1)}{\mathbb{Z}_6} = S(U(3)\times U(2))
\end{equation}  
and its (highly reducible) representation for the first generation of 16 basic fermions (and as many antifermions),  
 \begin{eqnarray}
 	\left( \begin{array}{c}
 		\nu       \\
 		e^-      \\
 	\end{array}\right)_L \leftrightarrow (\mathbf{1},\mathbf{2})_{-1}, \qquad
 	\left( \begin{array}{c}
 		u   \\
 		d   \\
 	\end{array}\right)_L\leftrightarrow  (\mathbf{3},\mathbf{2})_{\frac{1}{3}} \nonumber \\
 	 		(\nu_R \leftrightarrow (\mathbf{1},\mathbf{1})_0?)
 ,\qquad e^-_R \leftrightarrow(\mathbf{1}, \mathbf{1})_{-2}, \quad
 	u_R \leftrightarrow (\mathbf{3},\mathbf{1})_{\frac{4}{3}},\quad
 		d_R\leftrightarrow  (\mathbf{3},\mathbf{1})_{-\frac{2}{3}} 
 	\label{ba}
 \end{eqnarray}
  (the subscript standing for the value of the weak hypercharge $Y$),
  look rather baroque for a fundamental symmetry. Unsatisfied, the founding fathers 
  proposed Grand Unified Theories (GUTs) with (semi)simple symmetry groups: 
\smallskip
   $SU(5)$ \, H. Georgi - S.L. Glashow (1974);\\
   $Spin(10)$ \, H. Georgi (1975), \, H. Fritzsch - P. Minkowski (1975);\\
   $G_{PS}=Spin(6)\times Spin(4) = \frac{SU(4)\times SU(2) \times SU(2)}{\mathbb Z_2}$ \, J.C. Pati - A. Salam (1973). 
   
\smallskip   

The first two GUTs, based on simple groups, gained popularity in the beginning, since they naturally accommodated the fundamental fermions:     
\ba
\label{32}
SU(5): \mathbf{32} = \Lambda \mathbb{C}^5 = \bigoplus_{\nu=0}^5 \Lambda^{\nu},  \, \,
\Lambda^1=	\left( \begin{array}{c}
		\nu       \\
		e^-      \\
	\end{array}\right)_{-1}\oplus  
	\bar{d}_{\frac{2}{3}}= \mathbf{\bar{5}},   \nonumber\\    
	\Lambda^3 =	\left( \begin{array}{c}
u       \\
d       \\
\end{array}\right)_{\frac{1}{3}}\oplus 
\overline{u}_{-\frac{4}{3}}\oplus 
e^+_{2}= \mathbf{10}, \, \Lambda^5 = \bar{\nu}_L (?)=\bar{\textbf{•1}}; \, \nonumber\\
Spin(10): \mathbf{32}=\mathbf{16}_L \oplus \mathbf{16}_R, \, \mathbf{16}_L = \Lambda^1\oplus\Lambda^3\oplus\Lambda^5.
\ea 
(The question marks on the sterile (anti)neutrino indicate that their existence is only inferred indirectly - from the neutrino oscillations.) The Pati-Salam GUT is the only one to exploit the quark lepton symmetry: the group $SU(4)\subset G_{PS}$ combines the three colours with the lepton number. The left and right fermion octets are formed by $SU(2)_L$ and $SU(2)_R$ doublets, respectively (and conversely for the antifermions):
\be 
\label{PSfermios}
\textbf{8}_L=(\textbf{4}, \textbf{2}, \textbf{1}), \, \textbf{8}_R=(\bar{\textbf{•4}}, \textbf{1}, \textbf{2}) \, \,
(\bar{\textbf{8}}_R=(\bar{\textbf{•4}}, \textbf{2, \textbf{1)}}, \, \bar{\textbf{•8}}_L=(\textbf{4}, \textbf{1}, \textbf{2})).
\ee
The quark-lepton symmetry plays a pivotal role in our approach, too, and the Lie subalgebra $su(4) \oplus su(2)$ of 
$\mathfrak{g}_{PS}$ will appear in Sect. 4.1. 

If the fermions fit nicely in all GUTs, the gauge bosons start posing problems. The adjoint representations $\mathbf{24}$ (of $SU(5)$) and $\mathbf{45}$ (of $Spin(10)$) carry, besides the expected eight gluons and four electroweak gauge bosons, unwanted leptoquarks; for instance,
\begin{equation}
\label{24}
\mathbf{24}=(\mathbf{8}, \mathbf{1})_0\oplus (\mathbf{1}, \mathbf{3})_0\oplus (\mathbf{1},\mathbf{1})_0\oplus
(\mathbf{3}, \mathbf{2})_{\frac{5}{3}} \oplus (\mathbf{\bar{3}}, \mathbf{2})_{-\frac{5}{3}}.
\end{equation} 
Moreover, the presence of twelve gauge leptoquarks in (\ref{24}) yields a proton decay rate that contradicts current experimental bounds \cite{PDG114}. It is noteworthy that the Pati-Salam GUT is the only one which does not predict a gauge triggered proton decay (albeit it allows model dependent interactions with scalar fields that would permit such a decay).
Accordingly, the Pati-Salam group appears in a preferred reduction of the $Spin(10)$ GUT. Intriguingly, a version of this symmetry is also encountered in the noncommutative geometry approach to the SM, \cite{CCS}. Concerning the most popular nowadays \textit{supersymmetric GUTs} advocated authoritatively in \cite{W}, the lack of experimental evidence for any superpartner makes us share the misgivings expressed forcefully in \cite{Woi} (see also the recent popular account \cite{H}).

The noncommutative geometry approach, was started in 1988 (according to the dates of submission of \cite{DKM1, C}), see \cite{DKM, CL}, "at the height of the string revolution" (to cite \cite{CS19}) and pursued vigorously by Alain Connes and collaborators (work that can be traced back from \cite{CC, CCM, CCS, C17}) and by followers \cite{BF, L18} (for a pedagogical exposition see \cite{S}). 
 
The algebraic approach to quantum theory has, in fact, been initiated back in the 1930's by Pascual Jordan (1902-1980)\footnote{The only one of the "Boys' Club", \cite{J}, that did not get a Nobel Prize. The work on \textit{Jordan algebras} (called so by A.A. Albert, 1946),  of 1932-1934, that culminated in \cite{JvNW}, was preceded by the analysis (by Dirac, Jordan and von Neumann) of quantum transformation theory reviewed insightfully in \cite{DJ}. There are but a few papers on the applications of Jordan algebras to quantum theory, \cite{B93, B08, T85/16, GPR, DM, M, B12}.}, \cite{PJ}, who axiomatized the concept of \textit{observable algebra}, the prime example of which is the algebra of complex hermitian matrices (or self-adjoint operators in a Hilbert space) equipped with the symmetrized product
\begin{equation}
\label{circle}
A \circ B=\frac{1}{2}(AB+BA) \ (=B\circ A).
\end{equation}  
Such a (finite dimensional) Jordan algebra should appear as an "internal" counterpart of the algebra of smooth functions of classical fields. In the case of a \textit{special Jordan algebra} (i.e., a Jordan subalgebra of an associative algebra equipped with the product (\ref{circle})) one can of course work with its associative envelope, - i.e., with the corresponding matrix algebra. In the noncommutative geometry approach to the SM, based on a real spectral triple \cite{CC}, one arrives at the finite algebra \cite{CCM} (Proposition 3 of \cite{CS19}):
\begin{equation}
\label{NCGSM}
\mathcal{A}_F = \mathbb{C}\oplus\mathbb{H}\oplus\mathbb{C}[3]
\end{equation} 
($\mathbb{A}[n]$ standing for the algebra of $n\times n$ matrices with entries in the coordinate ring $\mathbb{A}$). The only hermitian elements of the quaternion algebra $\mathbb{H}$, however, are the real numbers, so $\mathcal{A}_F$ does not appear as the associative envelope of an interesting observable algebra. We shall, by contrast, base our treatment on an appropriate finite dimensional Jordan algebra\footnote{Recently, a new paper, \cite{BF19}, was posted where an alternative approach, closer to Connes' real spectral triple, involving a different Jordan algebra, is being developed.} suited for a quantum theory - permitting, in particular, a spectral decomposition  of observables.

\bigskip
 
\section{Euclidean Jordan algebras}
\setcounter{equation}{0}
\renewcommand\theequation{\thesection.\arabic{equation}}
An euclidean Jordan algebra is a real vector space $J$ equipped with a commutative product $X\circ Y$ with a unit 1 
satisfying the \textit{formal reality condition}
\begin{equation}
\label{formalR}
X_1^2+...+X_n^2 = 0 \Rightarrow X_1= ... =X_n=0 \,\, (X_i^2:=X_i\circ X_i)
\end{equation} 
and power associativity. Jordan has found a simple necessary and sufficient condition for power associativity. Introducing the operator $L(X)$ of multiplication by $X: L(X)Y=X\circ Y$, it can be written in the form:
\be
\label{LX}
[L(X), L(X^2)] =0 \Leftrightarrow X\circ (Y\circ X^2) = X^2\circ(Y\circ X), \, X, Y \in J.
\ee
(In general, non-associativity of the Jordan product is encoded in the noncommutativity of the maps $L(X)$.)
A prototype example of a Jordan algebra is the space of $n\times n$ hermitian matrices with anticommutator product (\ref{circle}), $X\circ Y = \frac{1}{2}(XY+YX)$ where $XY$ stands for the (associative) matrix multiplication. More generally, a Jordan algebra is called \textit{special} if it is a Jordan subalgebra of an associative algebra with Jordan product defined by (\ref{circle}). If $\mathcal{A}$ is an associative involutive  (star) algebra the symmetrized product  (\ref{circle}) is not the only one which preserves hermiticity. The quadratic (in $X$) operator $U(X) Y= XYX, X, Y\in \mathcal{A}$ also maps a pair $X, Y$ of hermitian elements  into a hermitian element. For a general (not necessarily special) Jordan algebra the map $U(X)$ (whose role is emphasized in \cite{McC}) and its polarized form $U(X, Y):= U(X+Y)-U(X)-U(Y)$ can be defined in terms of $L(X)$:
\be
\label{UL}
U(X) = 2 L^2(X) -L(X^2), U(X, Y) = 2(L(X)L(Y)+L(Y)L(X)-L(X\circ Y)).
\ee
  The conditions (\ref{formalR}) and (\ref{LX}) are necessary and sufficient to have spectral decomposition of any element of $J$ and thus treat it as an observable.

\subsection{Spectral decomposition, characteristic polynomial}
To begin with, power associativity  means that the subalgebra (including the unit) generated by an arbitrary element $X$ of $J$  is associative. In particular, any power of $X$  is defined unambiguously. In order to introduce spectral decomposition we need the algebraic counterpart of a projector. An element  $e\in J$ satisfying $e^2=e(\neq 0)$ is called a (non zero) \textit{idempotent}. Two idempotents $e$ and $f$ are \textit{orthogonal} if $e\circ f=0$; then multiplication by $e$ and $f$ commute and $e+f$ is another idempotent. The formal reality condition (\ref{formalR}) allows to define \textit{partial order} in $J$ saying that $X$ is smaller than $Y$, $X< Y$, if $Y-X$ can be written as a sum of squares. Noting that $f=f^2$ we conclude that $e<e+f$. A non-zero idempotent is called \textit{minimal} or \textit{primitive} if it cannot be decomposed into a sum of (nontrivial) orthogonal idempotents. A \textit{Jordan frame} is a set of orthogonal primitive idempotents $e_1, ..., e_r$ satisfying
\begin{equation}
\label{frame}
e_1+...+e_r=1 \, \,\, (e_i\circ e_j = \delta_{ij} e_i).
\end{equation} 
Each such frame gives rise to a complete set of commuting observables. The number of elements $r$ in a Jordan frame is independent of its choice and is called the \textit{rank of $J$}. Each $X\in J$ has a \textit{spectral decomposition} of the form
\begin{equation}
\label{Spec}
X= \sum_{i=1}^r \lambda_i e_i, \, \, \lambda_i\in \mathbb{R}, \, \, \lambda_1\leq\lambda_2\leq...\leq\lambda_r. 
\end{equation} 
For an $X$ for which all $\lambda_i$ in (\ref{Spec}) are different the spectral decomposition is unique. Such \textit{regular} $X$ form a dense open set in $J$. The rank of $J$ coincides with the degree of the characteristic polynomial (defined for any $X \in J$):
\ba
\label{CharPol}
F_r(t, X) =t^r-a_1(X)t^{r-1}+...+(-1)^r a_r(X), \nonumber \\ 
a_k(X)\in \mathbb{R}, \, a_k(\alpha X)=\alpha^k a_k(X) \, \, (\alpha>0). 
\ea
The roots of $F_r$ are $(t=)\lambda_1, ..., \lambda_r$ (some of which may coincide).
 Given a regular $X$ the idempotents $e_i$ can be expressed as polynomials in $X$ of degree $r-1$, determined from the system of equations 
\ba
\label{Vandermonde}
e_1+...+e_r=1, \nonumber \\
\lambda_1 e_1+...+\lambda_r e_r =X, \nonumber \\
 ... ...                            \nonumber \\
 \lambda_1^{r-1} e_1 +...+\lambda_r^{r-1} e_r = X^{r-1},
\ea
whose \textit{Vandermonde determinant} is non zero for distinct $\lambda_i$.

We are now ready to define a trace and an inner product in $J$.  The \textit{trace}, $tr(X)$, is a linear functional on $J$ taking value 1 on primitive idempotents: 
\begin{equation}
\label{trace}
tr(X) = \sum_i \lambda_i (=a_1(X)), \, \, \, tr(1)=r, 
\end{equation}
for $X$ given by (\ref{Spec}) (and $a_1(X)$ of  (\ref{CharPol})). The \textit{inner product}, defined as the trace of the Jordan  product, is positive definite:
\begin{equation}
\label{innerP}
(X, Y) := tr(X\circ Y) \Rightarrow (X, X) >0 \, \quad{for} \, X\neq 0.
\end{equation} 
This justifies the name \textit{euclidean} for a formally real Jordan algebra. The last coefficient, $a_r$, of  (\ref{CharPol}) is the \textit{determinant of $X$}:
\begin{equation}
\label{det}
a_r(X) = det(X)= \lambda_1...\lambda_r.  
\end{equation}
If $det(X)\neq 0$ then $X$ is \textit{invertible} and its inverse is given by
\begin{equation}
\label{inverse}
X^{-1}:=\frac{(-1)^r}{det(X)}(X^{r-1} -a_1(X)X^{r-1}+...+(-1)^{r-1}a_{r-1}(X) 1).
\end{equation}
 The theory of euclidean Jordan algebras
is simplified by the fact that any such algebra can be written as a direct sum of \textit{simple} ones (which cannot be further decomposed into nontrivial direct sums).

\subsection{Simple Jordan algebras. Euclidean extensions}
The finite dimensional simple euclidean Jordan algebras were classified at the dawn of the theory, in 1934, by Jordan et al.\footnote{A streamlined pedagogical version of the original argument, \cite{JvNW}, that follows \cite{K}, can be found in Chapters II-V of the book \cite{FK}.}. The argument is based on the \textit{Peirce decomposition} in a Jordan algebra which we are going to sketch.

To begin with, by repeated manipulation of the Jordan identity (\ref{LX}) one obtains the basic third degree formula (see Proposition II.1.1 (iii) of \cite{FK}):
\be\label{basic3}
L(X^2\circ Y)-L(X^2)L(Y) = 2(L(X\circ Y)-L(X)L(Y))L(X),
\ee
that is equivalent to
\ba 
L(X^3) - 3 L(X^2) L(X) + 2 L^3(X) = 0, \nonumber \\
\label{ThirdOrder}  [[L(X), L(Y)], L(Z)] + L([X, Z, Y]) = 0,  
\ea
$[X, Z, Y] := (X\circ Z)\circ Y - X\circ(Z\circ Y)$ is the \textit{associator}. For an idempotent, $X=e(= e^2)$, the first equation (\ref{ThirdOrder}) takes the form:
\be
\label{Le}
2L^3(e) -3L^2(e) + L(e) = L(e)(2L(e)-1)(L(e)-1) = 0,
\ee
thus restricting the eigenvalues of $L(e)$ to three possibilities $(0, 1/2, 1)$. 

Let $e\in J$ be a nontrivial idempotent: $0<e(=e^2)<1$. The eigensubspace $J_1(e)\subset J$ of $L(e)$ corresponding to eigenvalue 1 coincides with
$U(e)[J]$, the subspace of elements $Y$ of the form $Y=U(e)X,  X\in J$ where $U$ is the quadratic map (\ref{UL}). If the idempotent $e$ is minimal then $J_1(e)$ is one-dimensional: it us spanned by real mutiples of $e$. Similarly, the subspace $J_0(e)$ annihilated by 
$L(e)$ can be written as $J_0(e) = U(e^{'})[J]$ for. $e^{'} = 1-e$. For $J$ simple (of rank $r>1$) the subspace $J_{\frac{1}{2}}(e)$
has to be nontrivial as well:
\be
\label{J1/2}
J_{\frac{1}{2}}(e) = U(e, e^{'})[J] \Rightarrow (L(e)-\frac{1}{2}) J_{\frac{1}{2}}(e)= (L(e^{'})-\frac{1}{2}) J_{\frac{1}{2}}(e)=0.  
\ee

Given a frame of primitive idempotents $(e_1, ..., e_r)$ in a rank $r$ Jordan algebra $J_r$, we can introduce a set of $\binom{r+1}{2}$
orthogonal subspaces, a counterpart of Weyl's matrix units:
\be 
\label{JPWframe}
E_{ii}= U(e_i)[J_r], \, E_{ij}= U(e_i, e_j)[J_r](=E_{ji}) , \, \, i, j =1, ..., r, \, i\neq j.
\ee
They are eigenspaces of the set $\{L(e_k), k= 1, ..., r\}$ of commuting operators:
\be 
\label{eigen}
L(e_k) E_{ij} = \frac{1}{2}(\delta_{ik}+\delta_{jk}) E_{ij}.
\ee
The subalgebras $E_{ii}$ are one-dimensional while $E_{ij}, \, i\neq j$ (for a given simple algebra $J_r$) have the same dimension, called the degree, $d>0$. It turns out that the two positive numbers, the rank $r$ and the degree $d$, determine all finite dimensional simple euclidean Jordan algebras, to be, hence, denoted $J_r^d$. (The single rank one Jordan algebra is the field $\mathbb{R}$ of real numbers -  no room for off diagonal elements and no need for a degree in this case.) For $r=2$  the degree $d$ can take any positive integer value. For $r=3$ the allowed values of $d$ are the dimensions 1, 2, 4, 8 of the (normed) division rings. For $r\geq 4$ only the dimensions 1, 2, 4 of the associative division rings are permitted. The resulting simple Jordan algebras split into four infinite series and one exceptional algebra (proven to have no associative envelope by A. Albert also in 1934 and often called the \textit{Albert algebra}):
\begin{eqnarray}
J_r^1=\mathcal{H}_r(\mathbb{R}), \, r\geq 1; \quad J_r^2= \mathcal{H}_r(\mathbb{C}), \, r\geq 2; \nonumber \\
  J_r^4=\mathcal{H}_r(\mathbb{H}), \, r\geq 2; \quad \, \,  J_2^d= JSpin(d+1); \nonumber \\  
  J_3^8=\mathcal{H}_3(\mathbb{O}), \quad dim(J_r^d) = \binom{r}{2} d + r  
\label{ea1}
\end{eqnarray}
($dim(\mathcal{H}_r(\mathbb{R}))=\binom{r+1}{2}, \, dim(\mathcal{H}_r(\mathbb{C}))=r^2, \, dim(J_2^d)=d+2, \, dim(J_3^8)=27$).
The first three algebras in the above list consist of familiar  hermitian matrices (with entries in associative division rings).We stress once more that all items in (\ref{ea1}) (including $\mathcal{H}_r(\mathbb{C})$ and $\mathcal{H}_r(\mathbb{H})$ which involve matrices with complex and quaternionic entries) are regarded as algebras over the reals. The \textit{spin factor} $J_2^d\subset C\ell_{d+1}$ can be thought of as the set of $2\times 2$ matrices of the form
\ba
X=\xi \textbf{1}+\hat{x}, \, \xi\in\mathbb{R}, \,tr\hat{x}=0, \, X^2 = 2\xi X - det X, \nonumber \\
det X = \xi^2 - N(x), \,  \hat{x}^2=N(x)\textbf{1}, \, N(x)=\sum_{\mu=0}^d x_\mu^2  \label{JSpin}
\ea
(cf. Remark 3.1 of \cite{TD}). The fact that the algebras $JSpin(n)(=J_2^{n-1})$ are special requires an argument (while it is obvious for the first three series of matrix algebras (\ref{ea1})). In fact they admit interesting \textit{euclidean extensions}. 

The algebra $JSpin(n)$ is isomorphic to the Jordan subalgebra of the real Clifford algeb ra $C\ell_n$ spanned by the unit element and an orthonormal basis of gamma matrices with Jordan product
\be 
\label{Gamma(n)} 
\Gamma_i\circ \Gamma_j = \frac{1}{2} [\Gamma_i, \Gamma_j]_+ = \delta_{ij} \textbf{1}.
\ee
To define an appropriate euclidean extension we use the classification of real (later also of complex) Clifford algebras (see e.g. \cite{L}):
\ba 
C\ell_2=\mathbb{R}[2], \, C\ell_3= \mathbb{C}[2], \, C\ell_4 = \mathbb{H}[2], \, C\ell_5 = \mathbb{H}[2]\oplus \mathbb{H}[2], \,
C\ell_6 = \mathbb{H}[4], \,   \nonumber \\   C \ell_7 = \mathbb{C}[8], 
C\ell_8=\mathbb{R}[16], C\ell_9=\mathbb{R}[16]\oplus\mathbb{R}[16]; \, C\ell_{n+8} = C\ell_n \otimes \mathbb{R}[16]. \, \,  \label{CliffClass} 
\ea
It seems natural to define the euclidean extensions of the spin factors $JSpin(n)$ as the corresponding subalgebras of hermitian  (in the real case - symmetric) matrices: $\mathcal{H}_2(\mathbb{R}), \, \mathcal{H}_2(\mathbb{C}), \, \mathcal{H}_2(\mathbb{H}), ..., \, \mathcal{H}_{16}(\mathbb{R}) \oplus \mathcal{H}_{16}(\mathbb{R})$. In the case of real symmetric matrices (including the euclidean Jordan subalgebras of $C\ell_2, C\ell_8$ and $C\ell_9$), however, such a definition would exclude the most important obsevables: the hermitian counterparts of the symmetry generators. Indeed the derivations $\Gamma_{ab}=[\Gamma_a, \Gamma_b], a,b=1,...,n$ of $C\ell_n$ are antihermitian matrices; the hermitian observables $i\Gamma_{ab}$ only belong to the corresponding matrix Jordan algebra if we are dealing with complex hermitian (rather than real symmetric) matrices. More generally, we shall complexify from the outset the assoiciative envelope of the spin factors as postulated in \cite{D-V T}:
\ba 
J_2^d\in C\ell_{d+1}\in C\ell_{d+1}(\mathbb{C}), \, C\ell_{2m}(\mathbb{C})\cong \mathbb{C}[2^m], \nonumber \\ 
C\ell_{2m+1}(\mathbb{C})\cong \mathbb{C}[2^m]\oplus \mathbb{C}[2^m]. \, \,  \label{IfmplexEnv}
\ea
(See the insightful discussion in (Sect. 3 of) \cite{B12}.) We will identify the optimal extension $\tilde{J}_2^d$ of $J_2^d$ with the corresponding subalgebra of hermitian matrices. We shall exploit, in particular, 
\be 
\label{tildeJ}
\tilde{J}_2^8 := J_{16}^2\oplus J_{16}^2 = \mathcal{H}_{16}(\mathbb{C}) \oplus \mathcal{H}_{16}(\mathbb{C}).
\ee

Coming back to the list (\ref{ea1}) we observe that it involves three obvious repetitions: the spin factors $J_2^d$ for $d=1,2, 4$ coincide with the first items in the three families of matrix algebras in the above list. We could also write 
\begin{equation}
\label{JSpin9}
J_2^8 = \mathcal{H}_2(\mathbb{O}) (\subset C\ell_9);
\end{equation}
here (as in $J_3^8) \, \mathbb{O}$ stands for the nonassociative division ring of \textit{octonions} (see the review \cite{B}). The (10-dimensional) spin factor $J_2^8$ (unlike $J_3^8$) is \textit{special} - as a  Jordan subalgebra of the ($2^9$-dimensional) associative algebra $C\ell_9$.

\smallskip
\subsection{Symmetric cone, states, structure group}
Remarkably, an euclidean Jordan algebra gives room not only to the observables of a quantum theory, it also contains its states:  these are, roughly speaking, the positive observables. We proceed to more precise definitions. 

Each euclidean Jordan algebra $J$ contains a convex, \textit{open} \textit{cone} $\mathcal{C}$ consisting of all positive elements of $J$ (i.e., all invertible elements that can be written as sums of squares, so that all their eigenvalues are positive). Jordan frames belong to the closure $\bar{\mathcal{C}}$ (in fact, to the boundary) of the open cone, not to $\mathcal{C}$ itself, as primitive idempotents (for $r>1$) are not invertible. 

The \textit{states} are (normalized) positive linear functionals on the space of observables, so they belong to the closure of the dual 
cone
\begin{equation}
\label{dual}
\mathcal{C}^*= \{\rho\in J; (\rho, X)>0 \,\, \forall X\in \bar{\mathcal{C}}\}.
\end{equation} 
In fact, the positive cone is \textit{self-dual}, $\mathcal{C } = \mathcal{C }^*$. An element $\rho \in \bar{C}\subset J$ of trace one defines a \textit{state} assigning to any observable $X\in J$ an \textit{expectation value}
\begin{equation}
\label{expect}
<X> \, = (\rho, X) = tr(\rho\circ X), \, \rho\in\bar{\mathcal{C}}, \, \, tr\rho \, (=<1>)=1. 
\end{equation}
The primitive idempotents define \textit{pure states}; they are extreme points in the convex set of normalized states. All positive states (in the open cone $\mathcal{C}$) are (mixed) \textit{density matrices}. There is a distinguished mixed state in $J_r^d$, the normalized unit matrix, called by Baez the \textit{state of maximal ignorance}:
\begin{equation}
\label{unit}
<X>_0 = \frac{1}{r}tr(X) \, \, (r=tr(1)).
\end{equation} 
Any other state can be obtained by multiplying it by a (suitably normalized) observable - thus displaying a \textit{state observable correspondence} \cite{B12}.

The cone $\mathcal{C}$ is \textit{homogeneous}: it has a transitively acting symmetry group that defines the \textit{structure group} of the Jordan algebra, $Aut(\mathcal{C})=:Str(J)$, the product of a central subgroup $\mathbb{R}_+$ of uniform dilation with a (semi)simple Lie group $Str_0(J)$, the group that preserves the determinant of each element of $J$. Here is a list of the corresponding simple Lie algebras $str_0(J_r^d)$:
\ba
\label{str0}
str_0(J_r^1) = sl(r,\mathbb{R}), \, str_0(J_r^2) = sl(r, \mathbb{C}), \, str_0(J_r^4) = su^*(2r), \nonumber \\
str_0(J_2^d) = so(d+1, 1)(=spin(d+1, 1)), \, \, str_0(J_3^8) = \frak{e}_{6(-26)}.
\ea
The stabilizer of the point 1 of the cone is the maximal compact subgroup of $Aut(\mathcal{C})$ whose Lie algebra coincides with the derivation algebra\footnote{The corresponding automorphism group need not be connected. For instance, $Aut(J_r^2)$ is the semi-direct product of $SU(r)/\mathbb{Z}_r$ with a $\mathbb{Z}_2$ generated by complex conjugation.} of $J$:
\ba
\label{derJ}
\frak{der}(J_r^1) = so(r), \, \frak{der}(J_r^2) = su(r), \, \frak{der}(J_r^4) = usp(2r), \nonumber \\
\frak{der}(J_2^d) = so(d+1) (=spin(d+1)), \, \, \frak{der}(J_3^8) = \frak{f}_4.
\ea

The structure Lie algebra acts by automorphisms on $J$. For the simple Jordan algebras $J_r^d, \, d=1, 2, 4$ of hermitian matrices over an associative division ring an element $u$ of $str(J_r^d)$ transforms hermitian matrices into hermitian by the formula:
\be 
\label{uX}
u: (J_r^d \ni)X \rightarrow uX+Xu^*, \, \, d=1, 2, 4,
\ee
where $u^*$ is the hermitian conjugate of $u$. If $u$ belongs to the derivation subalgebra $\frak{der}(J_r^d)\subset str(J_r^d)$
then $u^*=-u$ and (\ref{uX}) becomes a commutator (thus annihilating the Jordan unit). In general, (\ref{uX}) can be viewed as a $Z_2$ graded commutator (regarding the hermitian matrices as odd elements). 

We shall argue that the exceptional Jordan algebra $J_3^8$ should belong to the observable algebra of the SM. It has three (special) Jordan subalgebras $J_2^8$ whose euclidean extensions match each one family of basic fermions.
 
\section{Octonions, quark-lepton symmetry, $J_3^8$}
\setcounter{equation}{0}
\renewcommand\theequation{\thesection.\arabic{equation}}
\subsection{Why octonions?}
The \textit{octonions} $\mathbb{O}$, the non-associative 8-dimensional composition algebra (reviewed in \cite{B, CS}, in Chapters 19, 23 of \cite{P95, L}, and in \cite{TD}, Sect. 1), were originally introduced as pairs of quaternions (the "Cayley-Dickson construction"). But it was the decomposition of $\mathbb{O}$ into complex spaces,
\ba
\mathbb{O} = \mathbb{C} \oplus \mathbb{C}^3, \, x=z+\mathbf{Z}, \, z=x^0+x^7e_7, \, \mathbf{Z} =Z^1e_1+Z^2e_2+Z^4e_4, 
 \nonumber \\  Z^j=x^j+x^{3j(mod 7)}e_7, \, j=1, 2, 4; \, \, \nonumber \\ 
 e_ie_{i+1}=e_{i+3(mod 7)}, \, \, e_ie_k + e_k e_i = -2\delta_{ik}, \, \, i, k = 1, ...7, \label{OC}
\ea
that led Feza G\"ursey (and his student Murat G\"unaydin) back in 1973,  \cite{GG, G}, to apply it to the quarks (then the newly proposed constituents of hadrons). They figured out that the subgoup $SU(3)$ of the automorphism group $G_2$ of the octonions, that fixes the first $\mathbb{C}$ in (\ref{OC}), can be identified with the quark colour group. G\"ursey tried to relate the non-associativity of the octonions to the quark confinement - the unobservability of free quarks. Only hesitantly did he propose (in \cite{G}, 1974, Sect. VII) "as another speculation" that the first $\mathbb{C}$ in (\ref{OC}) "could be related to leptons". 
Interpreting (\ref{OC}) as a manifestation of the quark-lepton symmetry was only taken seriously in 1987 by A. Govorkov \cite{Gov} in Dubna. The subject has been later pursued by G.M. Dixon, - see e.g. \cite{D10, D14} and in \cite{F16} among others. M. Dubois-Violette pointed out \cite{DV} that, conversely, the unimodularity of the quark's colour symmetry
 yields - through an associated invariant volume form - an essentially unique octonion product with a multiplicative norm. The octonions (just like the quaternions) do not represent an observable algebra. They take part, however, in $J_2^8$ and in the exceptional Jordan algebra $J_3^8$ whose elements obey the following Jordan product rules:
\ba 
\label{H3O}
X(\xi, x)& =&\left( \begin{array}{ccc}
	\xi_1  &  x_3  &  x_2^* \\
	x_3^* &  \xi_2 & x_1 \\
	x_2 & x_1^*  & \xi_3 \\
\end{array} \right) \nonumber\\
&=&\sum_{i=1}^3 (\xi_i E_i +F_i(x_i)), \, \, E_i\circ E_j = \delta_{ij}E_i, \, \, E_i\circ F_j = \frac{1-\delta_{ij}}{2} F_j, \nonumber \\
  F_i(x)\circ F_i(y) &=& (x, y) (E_{i+1}+E_{i+2}), \,
F_i(x) \circ F_{i+1}(y) =\frac{1}{2} F_{i+2}(y^* x^*)
\ea
(indices being counted mod 3). The (order three) characteristic equation for $X(\xi,x)$ has the form: 
\ba 
X^3 - tr(X) X^2 + S(X) X - det(X) = 0; \, \, tr(X) = \xi_1 + \xi_2 + \xi_3, \, \, \, \nonumber \\
 S(X)= \xi_1\xi_2 - x_3 x_3^* + \xi_2\xi_3- x_1x_1^* + \xi_1\xi_3 - x_2^*x_2, \, \, \nonumber \\
 det(X)= \xi_1\xi_2\xi_3 +2Re(x_1x_2x_3)-\sum_{i=1}^3 \xi_i x_ix_i^*. \label{CharX} 
\ea

The exceptional algebra $J_3^8$ incorporates  \textit{triality} that will be related - developing an idea of \cite{D-V T} - to the three generations of basic fermions. 
 
\smallskip

\subsection{Quark-lepton splitting of $J_3^8$ and its symmetry}
The automorphism group of $J_3^8$ is the compact exceptional Lie group $F_4$ of rank 4 and dimension 52 whose Lie algebra is spanned by the (maximal rank) subalgebra $so(9)$ and its spinorial representation $\textbf{16}$, and can be expressed in terms of $so(8)$ and its three (inequivalent) 8-dimensional representations:  
\begin{equation}
\label{f4}
\frak{der}(J_3^8) = \frak{f}_4 \cong so(9) + \mathbf{16} \cong so(8) \oplus \mathbf{8}_V \oplus \mathbf{8}_L \oplus \mathbf{8}_R;
\end{equation}
here $\mathbf{8}_V$ stands for the 8-vector, $\mathbf{8}_L$ and $\mathbf{8}_R$ for the left and right chiral $so(8)$ spinors. The group $F_4$ leaves the unit element $1=E_1+ E_2+ E_3$ invariant and transforms the traceless part of $J_3^8$ into itself (under its lowest dimensional fundamental representation $\textbf{26}$).  

The lepton-quark splitting (\ref{OC}) yields the following decomposition of $J_3^8$:
\ba
\label{H3z+Z}
X(\xi, x) = X(\xi, z) + Z, \, X(\xi, z)\in J_3^2 = \mathcal{H}_3(\mathbb{C}), \nonumber \\ 
Z=(Z^j_r, j=1, 2, 4, \, r=1, 2, 3)\in\mathbb{C}[3].
\ea
The subgroup of $Aut(J_3^8)$ which respects this decomposition is the commutant $F_4^\omega\subset F_4$ of the automorphism $\omega\in G_2\subset Spin(8)\subset F_4$ (of order three):
\begin{equation}
\label{omega}
\omega X(\xi, x)= \sum_{i=1}^3 (\xi_i E_i + F_i(\omega_7 x_i)), \, \, \omega_7=\frac{-1+\sqrt{3}e_7}{2} \, \, (\omega^3=1=
\omega_7^3).
\end{equation}
It consists of two $SU(3)$ factors (with their common centre acting trivially): 
\begin{equation}
\label{Sym3}
F_4^\omega = \frac{SU(3)_c\times SU(3)_{ew}}{\mathbb{Z}_3} \ni (U, V): \, X(\xi, x)\rightarrow VX(\xi, z)V^* + UZV^*.
\end{equation}
We see that the factor, $U$ acts on each quark's colour index $j(=1, 2, 4)$, so it corresponds to the exact $SU(3)_c$ colour symmetry while $V$ acts on the first term in (\ref{H3z+Z}) and on the flavour index $r(=1, 2, 3)$ and is identified with (an extension of) the broken electroweak symmetry as suggested by its restriction to the first generation of fermions (Sect. 4).

\smallskip

\section{The first generation algebra}
\setcounter{equation}{0}
\renewcommand\theequation{\thesection.\arabic{equation}}
\subsection{The $G_{SM}$ subgroup of $Spin(9)$ and observables in $\tilde{J}_2^8$}
The Jordan subalgebra $J_2^8\subset J_3^8$, orthogonal, say, to the projector $E_1$, 
\begin{equation}
\label{J1}
J_2^8(1) = (1-E_1)J_3^8(1-E_1),
\end{equation}
is special, its associative envelope being $C\ell_9$. Its automorphism group\footnote{Another instance of a (closed connected) maximal rank subgroup of a compact Lie group that have been studied by mathematicians \cite{BdS} back in 1949 (see also \cite{TD-V}). The group $Spin(9)$ had another appearance in the \textit{algebraic dreams} of P. Ramond \cite{R}.} is $Spin(9)\subset F_4$, whose intersection with $F_4^\omega$, that respects the quark-lepton splitting, coincides with - and \textit{explains} - \textit{the gauge group (1.1) of the SM}:
\begin{equation}
\label{GSM}
G_{SM}=F_4^\omega\cap Spin(9)= Spin(9)^\omega = S(U(3)\times U(2)).
\end{equation}
As argued in Sect. 2.2 the optimal euclidean extension $\tilde{J}_2^8$ (\ref{tildeJ}) of $J_2^8$ is obtained by replacing the real symmetric by complex hermitian matrices in the maximal euclidean Jordan subalgebra of its associative envelope:
\ba 
\label{EuclExt}
J_2^8\subset C\ell_9 = \mathbb{R}[16] \oplus \mathbb{R}[16] \rightarrow  J_{16}^1\oplus  J_{16}^1(\subset C\ell_9), \, 
J_{16}^1 = \mathcal{H}_{16}(\mathbb{R}) \nonumber \\
J_2^8 \rightarrow \tilde{J}_2^8 = J_{16}^2\oplus J_{16}^2 \subset C\ell_9(\mathbb{C}), \, \, \,  J_{16}^2 = \mathcal{H}_{16}(\mathbb{C}). \, \, \,
\ea
The resulting (reducible) Jordan algebra of rank 32 gives room precisely to the state space (of internal quantum numbers) of fundamental fermions of one generation - including the right handed "sterile" neutrino. In fact, it is acted upon by the simple structure group of $J_2^8$ whose generators belong to the even part of the Clifford algebra $C\ell(9,1)$ isomorphic to $C\ell_9$; its Dirac spinor representation splits into two chiral (Majorana-)Weyl spinors:
\begin{equation}
\label{StrJ}
Str_0(J_2^8) = Spin(9,1)\subset C\ell^0(9,1) (\cong C\ell_9)  \Rightarrow \textbf{32}=\textbf{16}_L \oplus \textbf{16}_R.
\end{equation}

Here is an explicit realization of the above embedding/extension. Let $e_\nu, \, \nu=0, 1, ...,7$, be the octonionic units satisfying (\ref{OC}). The anticommutation relations of the imaginary octonion units $e_j$ can be realized by the real skew-symmetric $8\times 8$ matrices $P_j$ generating $C\ell_{-6}$ and their product:
\be 
\label{CARofO}
[P_j, P_k]_+: = P_jP_k + P_kP_j=-2\delta_{jk}, P_7:=P_1...P_6\Rightarrow [P_7, P_j]_+=0, \, P_7^2=-1.
\ee  
The nine two-by-two hermitian traceless octonionic matrices $\hat{e}_a$ that generate $J_2^8$ (cf.(\ref{JSpin})) are represented by similar real symmetric $16\times 16$ matrices $\hat{P}_a$: 
\ba
\hat{e}_0=\sigma_1 e_0 \, (e_0=1), \hat{e}_j=c e_j, j=1, ..., 7, \, c=i\sigma_2, \, c^* = -c = c^3, \nonumber \\ 
 \hat{e}_a  \hat{e}_b +  \hat{e}_b  \hat{e}_a =2\delta_{ab}\Rightarrow \pm\hat{e}_8 = \hat{e}_0 ...\hat{e}_7 = \sigma_1 c^7 (-1) =-\sigma_3; \nonumber \\ \hat{P}_j= c\otimes P_j, j=1,...,7, \, \hat{P}_0= \sigma_1\otimes P_0, \, P_0=\textbf{1}_8, \, \hat{P}_8=  \sigma_3\otimes P_0. \label{ehat}
\ea
 The nine matrices $\hat{P}_a, a=0, 1, ..., 8$ generate an irreducible component of the Clifford algebra  $C\ell_9$. Then the ten real $32\times 32$ matrices
\ba  
\label{Gamm9-1}
\Gamma_a=\sigma_1 \otimes \hat{P}_a, \, a=0, 1, ..., 8, \, \Gamma_{-1}=c\otimes \textbf{1}_{16} \, \Leftrightarrow \, \nonumber \\ \Gamma_{-1}=\gamma_0 \otimes P_0, \Gamma_0=\gamma_1\otimes P_0, \Gamma_j=i\gamma_2\otimes P_j, j=1, ..., 7, \, 
\Gamma_8=\gamma_3\otimes P_0 \nonumber \\   
(\gamma_0= c^*\otimes \textbf{1}_2, \gamma_j = \sigma_1\otimes \sigma_j, j=1, 2, 3, \, \gamma_5=
i\gamma_0\gamma_1\gamma_2\gamma_3= \sigma_3 \otimes \textbf{1}_2)
\ea
generate the Clifford algebra $C\ell(9,1)$. We make correspond to the generators $\hat{e}_a$ (\ref{ehat}) of $J_2^8$ the hermitian elements $\Gamma_{-1}\Gamma_a$ of the even subalgebra $C\ell^0(9,1)\simeq C\ell_9$. The generators of the the symmetry algebra $so(9)$  of  the spin factor $J_2^8$ are given by the antihermitian matrices 
\be 
\label{GAMMAab} 
\Gamma_{ab}:= [\Gamma_{-1}\Gamma_a, \Gamma_{-1}\Gamma_b] = [\Gamma_a, \Gamma_b] (=-\Gamma_{ab}^*=.=-\Gamma_{ba}).
\ee
The corresponding observables $i\Gamma_{ab}$ only belong to the subalgebra $\tilde{J}_2^8$ (\ref{tildeJ}) of pairs of complex hermitian $16\times 16$ matrices.

In order to identify a complete commuting set of observables we introduce a maximal abelian subalgebra $\mathcal{A}$ of hermitian elements of the universal enveloping algebra $U(so(9,1))$ and the commutant $\mathcal{A}_8$ of $\Gamma_8$ in $\mathcal{A}$:
\ba 
\label{AA8}
\mathcal{A} = \mathbb{R}[\Gamma_{-1, 8}, i\Gamma_{07}, i\Gamma_{13}, i\Gamma_{26}, i\Gamma_{45}]\subset U(so(9,1)), \nonumber \\ 
\mathcal{A}_8=\mathbb{R}[i\Gamma_{07}, i\Gamma_{13}, i\Gamma_{26}, i\Gamma_{45}]\subset U(so(8)).
\ea
(The multilinear functions of $\mathcal{A}_8$ span a 16-dimensional vector subspace of the 32-dimensional real vector space 
$\mathcal{A}$.) To reveal the physical meaning of the generators of $\mathcal{A}_8\subset \mathcal{A}$ we identify them with the Cartan elements of the the maximal rank semisimple Lie subalgebra $so(6)\oplus so(3)$, the intersection 
\be 
\label{g4}
\mathfrak{g}_4:= so(6)\oplus so(3) \cong su(4)\oplus su(2) = \mathfrak{g}_{PS}\cap so(9)
\ee
 of the Pati-Salam algebra $\mathfrak{g}_{PS}=su(4)\oplus su(2)_L \oplus su(2)_R$ with $so(9)$, both viewed as Lie subalgebras of  $so(10)$. Here $su(2)$ is embedded  diagonally\footnote{We recall that while $I_3\in su(2)$ is a convenient global label for fundamental fermions, it is $su(2)_L\oplus u(1)$ which carries the local gauge symmetry of weakly interacting bosons.} into $su(2)_L\oplus su(2)_R$. It is easily verified that the Lie algebra $\mathfrak{g}_4$ acting on the vector representation  \textbf{9} of $Spin(9)$ preserves the quark lepton splitting in $J_2^8$. It does not preserve this splitting in the full Albert algebra $J_3^8$ which also involves the spinor representation \textbf{16} of $Spin(9)$. In accord with our statement in the beginning of this section  only its subgroup $G_{SM}$ (\ref{GSM}) does respect the required symmetry for both nontrivial IRs of $Spin(9)$ contained in the fundamental representation \textbf{26} of $F_4$.
 
  All elementary fermions can be labeled by the eigenvalues of two commuting operators in the centralizer of $su(3)_c$ in 
  $\mathfrak{g}_4(\subset so(9))$:
\be 
\label{I3B-L}
2I_3=-i\Gamma_{07}(=2I_3^L+2I_3^R), \, (2I_3)^2=1, \, \, B-L=\frac{i}{3}\sum_{j=1,2,4}\Gamma_{j, 3j(mod 7)}
\ee
and of the \textit{chirality} given by the Coxeter element of $C\ell(9,1)$:
\be 
\label{chirality}
\gamma:=\omega_{9,1} =\Gamma_{-1}\Gamma_0\Gamma_1...\Gamma_8=\gamma_5\otimes \textbf{1}_8\in C\ell^0(9,1) \, (=\gamma^*, \, \, \gamma^2 = 1).
\ee 
Here $B-L$ (the difference between the baryon and the lepton number) is given by the commutant of $su(3)_c$ in 
$so(6)(=Span\{\Gamma_{ik}, i, k =1, ...6\})\subset \mathfrak{g}_4$. The electric charge is a linear function of $B-L$ and $I_3$:
\be 
\label{QBL}
Q =\frac{1}{2}(B-L) + I_3.
\ee
The right and left isospins $I_3^{R/L}$ and the hypercharge $Y$ involve $\Gamma_{-1, 8}\notin \mathcal{A}_8$ and their expressions in terms of $I_3, B-L$ and $\gamma$ are more complicated:
\ba 
\label{I3RY}
4I_3^R = \Gamma_{-1, 8} -i\Gamma_{07}=(1+\gamma(B-L))(3\gamma(B-L)-1)(3\gamma(B-L)-2)I_3, \nonumber \\
4I_3^L = -\Gamma_{-1, 8} -i\Gamma_{07}=4(I_3 - I_3^R), \,  Y= B-L + 2I_3^R, \, \, \, \, \, \, \,
\ea
but we won't need them. Remarkably, all quarks and leptons are labeled by a single quantum number 
$B-L(=\pm 1, \pm 1/3)$ and two signs ($2I_3=\pm 1, \gamma =\pm$).


As the quark colour is not observable we only have to distinguish $SU(3)_c$ representations as labels: $\textbf{3}$ for a quark triplet, $\bar{\textbf{3}}$ for an antiquark, and $\textbf{1}$ for an $SU(3)_c$ singlet. These are encoded in the value of 
$B-L$: 
$$
B-L=\frac{1}{3}\leftrightarrow \textbf{3}, B-L=-\frac{1}{3}\leftrightarrow \bar{\textbf{3}}, B-L=\pm1\leftrightarrow \textbf{1}.
$$
 We have eight primitive idempotents corresponding to the left and right (anti)leptons and eight (non primitive) chiral (anti)quark idempotents (colour singlets of trace three); for instance,
\ba
\label{labels}
(\nu_L):=|\nu_L><\nu_L| \leftrightarrow ( 2I_3=1, B-L=-1, \gamma = 1),\, (e_R^+)\leftrightarrow (1, 1, -1); \nonumber \\ 
(\bar{u}_L):=\sum_{j=1,2,4}|\bar{u}_L^j>< \bar{u}_L^j|\leftrightarrow (-1, -\frac{1}{3}, 1), \,  (d_L) \leftrightarrow 
(-1, \frac{1}{3}, 1). \, \, \,
\ea 

\smallskip

\subsection{Observables. Odd chirality operators}
The 16-dimensional vector spaces $S_R$ and $S_L$ can be identified with the subspaces of $J_3^8$ spanned by $F_2(x_2)$ and
$F_3(x_3)$ of Eq. (\ref{H3O}), respectively. (As we shall recall in Sect. 5.1 below, they transform as expected under the $Spin(8)$ subgroup of the automorphism group $F_4$ of $J_3^8$.) The (extended) observables belong by definition to the complexification of the even part, $C\ell^0(9,1)$, of $C\ell(9,1)(\cong \mathbb{R}[32])$, thus commute with chirality 
and preserve individually the spaces $S_R$ and $S_L$. The elements of the odd subspace, $C\ell^1(9,1)$, in particular the generators $\Gamma_\mu$ of $C\ell(9,1)$, by contrast, anticommute with chirality and interchange the left and right spinors. They can serve to define the internal space part of the Dirac operator.

We shall now present an explicit realization of both $\Gamma_\mu$ and the basic observables in terms of \textit{fermionic oscillators} (anticommuting creation and annihilation operators - updating Sect. 5 of \cite{TD}), inspired by \cite{A95, F18}. 
To begin with, we note that the complexification of $C\ell(9,1)$ contains a five dimensional isotropic subspace spanned by the anticommuting operators
\ba 
a_0=\frac{1}{2}(\Gamma_0+i\Gamma_7), a_1=\frac{1}{2}(\Gamma_1+i\Gamma_3), a_2=\frac{1}{2}(\Gamma_2+i\Gamma_6), a_4 = \frac{1}{2}(\Gamma_4+i\Gamma_5)  \nonumber \\ 
(i.e. \, a_j=\frac{1}{2}(\Gamma_j+i\Gamma_{3j mod7}), \, j=1, 2, 4, \,  i^2=-1), \, a_8 = \frac{1}{2}(\Gamma_8 +\Gamma_{-1}). \, 
\, \, \, \label{fermion} 
\ea
The $a_\mu$ and their conjugate $a_\mu^*$ obey the canonical anticommuttation relations,
\be
\label{CAR} [a_\mu, a_\nu]_+=0, \, [a_\mu, a_\nu^*]_+=\delta_{\mu\nu}, \, \mu, \nu = 0, 1, 2, 4, 8,
\ee
equivalent to the defining Clifford algebra relations for $\Gamma_\alpha, \, \alpha=-1, 0, 1, ..., 8$. 

\textit{Remark}. The $32\times 32$ matrices $\Gamma_{-1}, \Gamma_\mu (\mu=0, 1, 2, 4, 8), i\Gamma_7, i\Gamma_3, i\Gamma_6, 
i\Gamma_5$ generate the split real form $C\ell(5, 5)(\cong C\ell(9,1))$ of $C\ell_{10}(\mathbb{C})$. The split forms $C\ell(n, n)$ (which allow to treat spinors as differential forms) have been used by K. Krasnov \cite{K18} (for $n=7$) in his attempt to make the SM natural.  

We now proceed to translate the identification of basic observables of Sect.4.1 in terms of the five products $a_\mu^*a_\mu, \, \mu=0, 1, 2, 4, 8$. The set of  15 quadratic combinations of $a_i^*, a_j$ (nine $a_i^* a_j$, three independent products $a_i a_j=-a_j a_i$, and as many $a_i^* a_j^*$) generate the Pati-Salam Lie algebra $su(4)$. The centralizer $B-L$ (\ref{I3B-L}) of $su(3)_c$ in $su(4)$ now assumes the form:
\be 
\label{B-L}
B-L = \frac{1}{3}\sum_j [a_j^*, a_j] \Rightarrow [B-L, a_j^*] = \frac{1}{3}a_j^*, \, 
 [B-L, a_1a_2a_4] = -a_1a_2a_4
\ee
($j=1, 2, 4$). The indices $0, 8$ correspond to the left and right isospins:
\ba 
\label{IL&IR}
I_+^L=a_8^*a_0, \, I_-^L=a_0^*a_8,  \, 2I_3^L=[I_+^L, I_-^L] = a_8^* a_8 - a_0^*a_0; \nonumber \\
I_-^R=a_0^*a_8^*, \, I_+^R=a_8a_0, \, 2I_3^R=[I_+^R, I_-^R] =a_8a_8^*-a_0^*a_0; \nonumber \\
I_+:= I_+^L+I_+^R=\Gamma_8 a_0, \, I_-:= I_-^L+I_-^R= a_0^*\Gamma_8, \, 2I_3=[I_+, I_-]=[a_0,a_0^*]. \, \,
\ea
The hypercharge and the chirality also involve $\Gamma_{-1}\Gamma_8=[a_8, a_8^*]$:
\be 
\label{YQ} Y=\frac{2}{3}\sum_j a_j^*a_j -a_0^*a_0 -a_8^*a_8 , \, \, \gamma=[a_0^*, a_0][a_1^*, a_1][a_2^*, a_2][a_4^*, a_4]
[a_8^*, a_8].
\ee

All fermion states and the commuting observables are expressed in terms of five basic idempotents $\pi_\mu$ and their complements:
\ba 
\pi_\mu = a_\mu a_\mu^*, \, \bar{\pi}_\mu=1-\pi_\mu = a_\mu^*a_\mu, \pi_\mu^2=\pi_\mu, \, \bar{\pi}_\mu^2 = \bar{\pi}_\mu, 
\pi_\mu\bar{\pi}_\mu =0, \, [\pi_\mu, \pi_\nu] = 0, \nonumber \\
\label{basicIdem} \pi_\mu + \bar{\pi}_\mu = 1, \, tr 1 = 32 \Rightarrow tr\pi_\mu=tr\bar{\pi}_\mu =16, \, \mu=0, 1, 2, 4, 8. \, \, \, \, \, \,
\ea
As $\pi_\mu$ commute among themselves, products of $\pi_\mu$ are again idempotents. No product of less than five factors is primitive; for instance $\pi_0\pi_1\pi_2\pi_4= \pi_0\pi_1\pi_2\pi_4(\pi_8+\bar{\pi}_8)$. Any of the $2^5$ products of five  
$\pi_\mu, \bar{\pi}_\nu$ of different indices is primitive. We shall take as "vacuum" vector the product of all $\pi_\mu$
which carry the quantum numbers of the right chiral ("sterile") neutrino:
\be 
\label{Omega}
\Omega := \pi_0\pi_1\pi_2\pi_4\pi_8 = (\nu_R) \leftrightarrow (2I_3=1, B-L =-1, \gamma=-1), \, a_\nu\Omega=0.
\ee
The left chiral states involve products of an even number of $\pi_\nu$ (an odd number for the right chiral states). In particular, $(\bar{\nu}_L)$, the antiparticle of $(\nu_R)$, only involves $\bar{\pi}_\nu$ factors:
\be 
\label{barOmega}
 (\bar{\nu}_L)=\bar{\Omega}:= \bar{\pi}_0\bar{\pi}_1\bar{\pi}_2\bar{\pi}_4\bar{\pi}_8\leftrightarrow (-1, 1, 1), \, a_\nu^*\bar{\Omega}=0.
\ee 

Each primitive idempotent can be obtained from either $\Omega$ or $\bar{\Omega}$ by consecutive action of the involutive chirality changing bilinear maps (cf. (\ref{UL})):
\be 
\label{odd inv} 
U(a_\mu^*,a_\mu)X=a_\mu^* X a_\mu + a_\mu Xa_\mu^*, \, X\in \mathcal{A}=\mathbb{R}[\pi], \, \mu=0, 1,2, 4, 8;
\ee 
here $\pi=\{\pi_0, \pi_1, \pi_2, \pi_4, \pi_8\}$ is a basis of idempotents of the abelian multilinear algebra $\mathcal{A}$ 
(\ref{AA8}). More economically, we obtain all eight lepton states by acting with the above operators for $\mu=0, 8$ on both $\Omega$ and  $\bar{\Omega}$ (or only on $\Omega$ but also using the "colourless" operator $U(a_1^*a_2^*a_4^*, a_4a_2a_1)$):
\ba 
(\nu_R)=\Omega, \, (\nu_L)=U(a_0^*,a_0)\Omega=a_0^*\Omega a_0, \, (e_L^-)=a_8^*\Omega a_8 \, (a_0\Omega a_0^*=0=a_8\Omega a_8^*); \nonumber \\  
(e_R^-)=a_0^*a_8^*\Omega a_8a_0, \, \, (e_L^+)=a_8a_0\bar{\Omega}a_0^*a_8^*=a_1^*a_2^*a_4^*\Omega a_4a_2a_1; \nonumber \\
\label{leptons}  (e_R^+)=a_8\bar{\Omega}a_8^*=a_0^*(e_L^+)a_0, \,
(\bar{\nu}_R)=a_0\bar{\Omega}a_0^*=a_8^*(e_L^+)a_8, \, (\bar{\nu}_L)=\bar{\Omega}=a_0^*(\bar{\nu}_R)a_0. \, \, \, \, \,\,
\ea
The $SU(3)_c$ invariant (trivalent) quark states are obtained by acting with the operator
\be 
\label{Uq}
 U_q:=\sum_{j=1,2,4} U(a_j^*, a_j)
\ee
on the corresponding lepton states:
\ba 
U_q(\nu_R) = (\bar{d}_L), \, U_q(e_R^-)=(\bar{u}_L); \, \, U_q(\nu_L)=(\bar{d}_R), \ U_q(e_L^-)=(\bar{u}_R); \nonumber \\
U_q(e_L^+)=(u_R), \, U_q(\bar{\nu}_L)=(d_R); \, \, U_q(e_R^+)=(u_L), \, U_q(\bar{\nu}_R)=(d_L). \label{lepton-quark} 
\ea

We note that while $(\nu_R)$ is the lowest weight vector of $so(9,1)$ with five simple roots corresponding to the commutators
$[a_\mu^*, a_\mu]$,
the state that minimizes our choice of observables, $2I_3, B -L, \gamma$, is the right electron singlet $(e_R^-)$ - the lowest weight vector of $\mathfrak{g}_4\times \gamma$. (The corresponding highest weight vectors are given by the respective antiparticle states.)
 
\smallskip\

\textit{Remark}. Each of the rank 16 extensions of $J_2^8$ appears as a subrepresentation of the defining module $\textbf{26}$ of the automorphism group $F_4$ of $J_3^8$, which splits into three irreducible components of $Spin(9)$:
\be 
\label{26=16+10}  
 \textbf{26} = \textbf{16 + \textbf{9} +\textbf{1}}.
 \ee
 The vector representation $\textbf{9}$ of $SO(9)$ is spanned by the generators $\Gamma_a$ of $C\ell_9$. Their splitting into the defining representations of the two factors of its maximal rank subgroup $SO(6)\times SO(3)$ displays the quantum numbers of a pair of conjugate leptoquark and a triplet of weak interaction bosons: 
 $$\textbf{9} = (\textbf{6}, \textbf{1})\oplus (\textbf{1}, \textbf{3}); \, \,  \textbf{6} = (a_j, a_j^*, \, j=1, 2, 4) =(D, \bar{D}); $$ 
$$[B-L,a_j]= -\frac{2}{3}a_j, \, [B-L, a_j^*] = \frac{2}{3}a_j^* 
\Rightarrow 
D\leftrightarrow (0, -\frac{1}{3}, -\frac{2}{3}), \, \bar{D}\leftrightarrow (0, \frac{1}{3}, \frac{2}{3});$$
\begin{eqnarray}
\textbf{3}=(a_0^*, \Gamma_8, a_0)=:(W^+, W^0, W^-)\leftrightarrow Y=B-L=0, \, Q=(1,0,-1). \, \, \label{DW} 
\end{eqnarray}
Combining the $\textbf{3}$ with the singlet $\textbf{1}$ one can find the mixtures that define the (neutral) Z-boson and the photon. The leptoquarks $D, \bar{D}$ also appear in the $\textbf{27}$ of the $E_6$ GUT (see, e.g. \cite{Sch} where they are treated as superheavy).

As pointed out in the beginning of this subsection we shall view instead the $\textbf{9}\oplus \textbf{1}$ as the gamma matrices which anticommute with chirality and (are not observables but) should enter the definition of the Dirac operator.

\smallskip
 
\section{Triality and Yukawa coupling}
\setcounter{equation}{0}
\renewcommand\theequation{\thesection.\arabic{equation}}

\subsection{Associative trilinear form. The principle of triality}
The trace of an octonion $x=\sum_\mu x^\mu e_\mu$ is a real valued linear form on $\mathbb{O}$:
\begin{equation}
\label{trace}
tr(x) = x +x^* = 2x^0 =2Re(x) \, \, (e_0\equiv 1).
\end{equation}
It allows to define an associative and symmetric under cyclic permutations \textit{normed triality} form $t(x, y, z)= Re(xyz)$ satisfying:
\begin{equation}
\label{t8}
2t(x,y,z)=tr((xy)z)=tr(x(yz))=:tr(xyz) = tr(zxy)=tr(yzx).
\end{equation} 
The normalization factor 2 is chosen to have:
\begin{equation}
\label{norm}
|t(x,y,z)|^2 \leq N(x)N(y)N(z), \,\, N(x)=xx^*(\in \mathbb{R}).
\end{equation}
While the norm $N(x)$ and the corresponding scalar product are $SO(8)$-invariant, the trilinear form $t$ corresponds to the invariant product of the three inequivalent 8-dimensional fundamental representations of $Spin(8)$, the 8-vector and the two chiral spinors, say $S^\pm$ (denoted by $\textbf{8}_L$ and $\textbf{8}_R$ in (\ref{f4})).  We proceed to formulating the more subtle trilinear invariance law.\footnote{For systematic expositions of the $Spin(8)$ triality see, in order of appearance, \cite{P95} (Chapter 24), \cite{L} (Chapter 23), \cite{B} (Sect. 2.4), \cite{CS} (Sect. 8.3), \cite{Y} (Sects. 1.14-1.16).} 

\textbf{Theorem 5.1} (\textit{Principle of triality} - see \cite{Y}, Theorem 1.14.2). For any $g\in SO(8)$ there exists a pair ($g^+, g^-$) of elements of $SO(8)$, such that
\begin{equation}
\label{PrincTri}
g(xy) = (g^+x)(g^-y), \,\, x, y\in \mathbb{O}.
\end{equation} 
If the pair $(g^+, g^-)$ satisfies (\ref{PrincTri}) then the only other pair which obeys the principle of triality is $(-g^+, -g^-)$.

\textit{Corollary}. If the triple $g, g^+, g^-$ obeys (\ref{PrincTri}) then the form $t$ (\ref{t8}) satisfies the invariance condition
\begin{equation}
\label{t-inv}
t(g^+x, g^-y, g^{-1}z) = t(x, y, z).
\end{equation}

\textbf{Proposition 5.2} (see \cite{Y} Theorem 1.16.2). The set of triples $(g, g^+,, g^-)\in SO(8)\times SO(8)\times SO(8)$
satisfying the principle of triality form a group isomorphic to the double cover $Spin(8)$ of $SO(8)$.

An example of a triple $(g^+, g^-, g^{-1})$ satisfying (\ref{t-inv}) is provided by left-, right- and bi-multiplication by a unit octonion:
\begin{equation}
\label{Lu}
t(L_ux, R_uy, B_{u^*}z) = t(x, y, z), \, L_ux = ux, R_uy=yu, B_vz=vzv, \, \, uu^*=1.
\end{equation}
The permutations among $g, g^+, g^-$ belong to the group of outer automorphisms of the Lie algebra $so(8)$ which coincides with the symmetric group $\mathcal{S}_3$ that permutes the nodes of the Dynkin diagram for $so(8)$. In particular, the map that permutes cyclicly $L_u, R_u, B_{u^*}$ belongs to the subgroup $\mathbb{Z}_3\subset \mathcal{S}_3$:
\begin{equation}
\label{Z3}
\nu : L_u \rightarrow R_u\rightarrow B_{u^*} \Rightarrow \nu^3 = 1.
\end{equation} 

\textit{Remark.}  The associativity law expressed in terms of left (or right) multiplication reads
\begin{equation}
\label{assoc}
L_x L_y = L_{xy}, \, R_x R_y = R_{yx}.
\end{equation} 
It is valid for complex numbers and for quaternions; for octonions Eq. (\ref{assoc}) only takes place for real multiples of powers of a single element. Left and right multiplications by unit quaternions generate different $SO(3)$ subgoups of the full isometry group $SO(4)$ of quaternions. By contrast, products of upto 7 left multiplications of unit octonions (and similarly of upto 7 $R_u$ or $B_u$) generate the entire $SO(8)$ (see Sect. 8.4 of \cite{CS}).

\smallskip

\textbf{Proposition 5.3} (see \cite{Y} Theorem 2.7.1). The subgroup $Spin(8)$ of $F_4$ leaves the diagonal projectors $E_i$ in the generic element $X(\xi, x)$ (\ref{H3O}) of $J_3^8$ invariant and transforms the off diagonal elements as follows:
\begin{equation}
\label{FiSpin8}
F_1(x_1)+ F_2(x_2) + F_3(x_3) \rightarrow F_1(gx_1) +F_2(g^+x_2)+ F_3(g^-x_3).
\end{equation}

Thus if we regard $x_1$ as a $Spin(8)$ vector, then $x_2$ and $x_3$ should transform as $S^+$ and $S^-$ spinors, respectively.

\smallskip

\subsection{Speculations about Yukawa    couplings}

It would be attractive to interpret the invariant trilinear form $t(x_1, x_2, x_3)$ as the internal symmetry counterpart of the Yukawa coupling between a vector and two (conjugate) spinors. Viewing the 8-vector $x_1$ as the finite geometry image of the Higgs boson, the associated Yukawa coupling would be responsible for the appearance of (the first generation) fermion masses. 
There are, in fact, three possible choices for the $SO(8)$ vector, one for each generation $i$ corresponding to the Jordan subalgebra
\begin{equation}
\label{3gen}
J_2^8(i) = (1-E_i)J_3^8(1-E_i), \, i=1, 2, 3.
\end{equation} 
According to Jacobson \cite{J68} any finite (unital) module over $J_3^8$ has the form $J_3^8 \otimes E$ for some finite dimensional real vector space $E$. The above consideration suggests that $dim(E)$ should be a multiple of three so that there would be room for an octonion counterpart of a vector current for each generation.

\smallskip 

As demonstrated in Sect. 4 the natural euclidean extension of $J_2^8$ gives rise to a Jordan frame fitting nicely one generation of fermions. We also observed that the generators $\Gamma_a$ of $C\ell_9$ anticommute with chirality and thus do not belong to the observable algebra but may serve to define (the internal part of) the Dirac operator. Unfortunately, according to Albert's theorem, $J_3^8$ admits no associative envelope and hence no such an euclidean extension either. To search for a possible substitute it would be instructive to see what exactly would go wrong if we try to imitate the construction of Sect. 4.1. The first step, the map (cf. (\ref{CARofO}))
\ba 
\mathbb{O}\ni x\rightarrow P(x)=\sum_{\alpha=0}^7 x^\alpha P_\alpha \in C\ell_{-6}\cong \mathbb{R}[8], \nonumber \\
P(xx^*) = P(x)P(x)^* = xx^* P_0, \, P(x)^*=\sum P_\alpha^* x^\alpha, \, P_0=\textbf{1}_8, \label{mapOCl}
\ea
for each of the arguments of $F_i, i=1, 2, 3$, respects all binary relations. We have, for instance (in the notation of (\ref{H3O})),
\be 
F_1(P(x_1))\circ F_2(P(x_2))=\frac{1}{2}F_3(P(x_2^*x_1^*)), \, P(x^*y^*)=P(x)^* P(y)^* \, .
\label{FPx}
\ee
The map (\ref{mapOCl}) fails, however, to preserve the trilinear form (\ref{t8}) which appears in $det(X)$ (\ref{CharX}) and in triple products like
\be 
\label{F(x)}
(F(x_1)\circ F(x_2))\circ F(x_3) = \frac{1}{2}t(x_1,x_2,x_3) (E_1 +E_2); 
\ee
setting $P(x_i)=X_i$ we find instead of $t$ more general $8\times 8$ matrices:
\ba 
(F_1(X_1)\circ F_2(X_2))\circ F_3(X_3) =  \nonumber \\
\frac{1}{4}((X_2^*X_1^*X_3^* + X_3X_1X_2) E_1 + (X_1X_2X_3+X_3^*X_1^*X_2^*)E_2) \label{F123}.
\ea
Moreover, while $t(e_1, e_2, e_3)=0, t(e_1, e_2, e_4)=-1$, the corresponding (symmetric, traceless, mutually orthogonal) matrices $P_1P_2P_3, P_1P_2P_4$ have both square 1 and can be interchanged by an inner automorphism.  

In order to apply properly the full exceptional Jordan algebra to particle dynamics we need further study of differential calculus and connection forms on Jordan modules, as pursued in \cite{CDD, DV18}, on one side, and connect with current phenomenological understanding of the SM, on the other. In particular, the Cabibbo-Kobayashi-Maskawa (CKM) quark- and the Pontecorvo-Maki-Nakagawa-Sakata (PMNS) lepton-mixing matrices \cite{PDG12, PDG14} should be reflected in a corresponding mixing of their finite geometry counterpart. Its concrete realization should help us fit together the euclidean extensions of the three subalgebras $J_2^8(i)$ and the Yukawa couplings expressed in terms of the invariant trilinear form in $J_3^8$.


\smallskip

\section{Outlook}
\setcounter{equation}{0}
The idea of a finite quantum geometry appeared in the late 1980's in an attempt to make the Standard Model natural, avoiding at the same time the excessive number of new unobserved states accompanying GUTs and higher dimensional theories. It was developed and attained maturity during the past thirty years in the framework of noncommutative geometry and the real spectral triple in  work of Alain Connes and others (\cite{C17, BF, CS19}). Here we survey the progress in an alternative attempt, put forward by Michel Dubois-Violette, \cite{DV} (one of the originators of the noncommutattive geometry approach, too) based on a finite dimensional counterpart of the \textit{algebra of observables}, hence, a (commutative) euclidean Jordan algebra. As the deep ideas of Pascual Jordan seem to have found a more receptive audience among mathematicians than among physicists, we used the opportunity to emphasize (in Sect. 2) how well suited a Jordan algebra $J$ is for describing both the observables and the states of a quantum system.  

 Recalling (Sect. 2.2) the classification of finite dimensional simple euclidean observable algebras (of \cite{JvNW}, 1934) and the quark-lepton symmetry (Sect. 3.1) we argue that it is a multiple of the exceptional Jordan (or \textit{Albert}) algebra $J_3^8$ that describes the three generations of fundamental fermions.  We postulate that the symmetry group of the SM is the subgroup $F_4^\omega$ of $Aut(J_3^8)=F_4$ that respects the quark-lepton splitting. Remarkably, the intersection of $F_4^\omega$ with the automorphism group $Spin(9)$ of the subalgebra $J_2^8\subset J_3^8$ of a single generation is just the gauge group of the SM.   

The  next big problem we should face is to fix an appropriate $J_3^8$ module (following our discussion in Sect. 5.2) and to write down the Lagrangian  in terms of fields taking values in this module.

\smallskip 

\footnotesize{This paper can be viewed as a progress report on an ongoing project pursued in \cite{DV, TD-V, TD, D-V T}.  I thank my coauthors in these papers and Cohl Furey for discussions, Latham Boyle and Svetla Drenska for a careful reading of the manuscript, and both her and Michail Stoilov for their help in preparing these notes. I am grateful to Vladimir Dobrev, Latham Boyle, George Zoupanos and Andrzej Sitarz for invitation and hospitality in Varna, at the Perimeter Institute, in Corfu and in Krak\'ow, respectively.}   
\bigskip


\end{document}